\documentclass[12pt,a4paper,dvips]{article}
\usepackage{a4p}
\usepackage{cite,mcite}
\usepackage{graphicx}
\usepackage{subfigure,rotating}
\usepackage{physics}
\usepackage{l3_title,ifthen}
\usepackage{Lep}
\usepackage{amssymb,amsmath}
\usepackage{epsfig}
\journalname{Phys. Lett. B}
\date{May 04, 2000}
\preprint{2000-064}
\Lep{2}
\newlength{\capindent}
\newlength{\capwidth}
\newlength{\figwidth}
\newcommand{\icaption}[2][!*!,!]{\hspace*{\capindent}%
  \begin{minipage}{\capwidth}
    \ifthenelse{\equal{#1}{!*!,!}}%
      {\caption{#2}}%
      {\caption[#1]{#2}}
  \end{minipage}}
\def\as{\alpha_{\mathrm{s}}}

\def\bt{B_{T}}
\def\bw{B_{W}}
\def\ee{\mathrm{e^{+}e^{-}}}

\def\epar{E_{\parallel}}
\def\eperp{E_{\perp}}
\def\evis{E_{\mathrm{vis}}}
\def\WW{\mathrm{W^{+}W^{-}}}

\def\rs{\sqrt{s}}

\def\qq{\mathrm{q\bar{q}}}

\def\Nf{{N_{f}}}
\def\Nc{{N_{c}}}

\def\pb{\mathrm{pb^{-1}}}

\begin{document}
\bibliographystyle{l3style}
\begin{titlepage}
\date{\today}

\title{QCD Studies in {\boldmath $\ee$} Annihilation \\
       from 30 \GeV\ to 189 \GeV}
\author{The L3 Collaboration}
%
%
\begin{abstract}
We present results obtained from a study of the structure of hadronic 
events recorded by the L3 detector at various centre-of-mass energies.
The distributions of event shape variables and the energy dependence of 
their mean values are measured from 30 \GeV\ to 189 \GeV\  and compared with 
various QCD models. The energy dependence of the moments of event shape 
variables is used to test a power law ansatz for the non-perturbative
component. We obtain a universal value of the non-perturbative parameter
$\alpha_{0}$ = 0.537 $\pm$ 0.073. From a comparison 
with resummed $\cal{O}$$(\as^{2})$ QCD calculations, we determine the 
strong coupling constant at each of the selected energies. The measurements 
demonstrate the running of $\as$ as expected in QCD with a value of  
$\as (\MZ)$ = 0.1215 $\pm$ 0.0012 (exp) $\pm$ 0.0061 (th).
\end{abstract}
\submitted

\end{titlepage}

\normalsize

\section{Introduction}\label{sec:intr}

LEP operated at centre-of-mass energies around 91.2 \GeV\  from 1989 to 
1995 and then moved up to six different 
centre-of-mass energies between 130 \GeV\ and 189 \GeV\  
in the following three years. 
Thus a study of the process $\ee$ $\rightarrow$ hadrons at LEP 
offers a unique environment to test the predictions of the theory of the 
strong interaction (QCD) over a wide energy range. The energy range
has been extended by using hadronic events from Z decays with isolated 
high energy photons in order to probe the structure of hadronic
events at reduced centre-of-mass energies down to 30 GeV \cite{l3qcd91,l3qqg}. 
The high energy photons are radiated early in the process through 
initial state radiation (ISR) or through quark bremsstrahlung whereas the 
hadronic shower develops over a longer time scale.

We report here measurements of event shape distributions and their 
moments using the data collected with the L3 detector~\cite{l3:det}. 
We update the published results at $\rs$ = 161, 172 and 183 \GeV\ 
\cite{l3qcd133,l3qcdlep2} with an improved selection method for hadronic 
events and present new results at $\rs$ = 130, 136 and 189 \GeV. 
The measured distributions are compared with predictions from 
event generators based on an improved leading log approximation (Parton 
Shower models including QCD coherence effects). Three such Monte Carlo 
programs (\textsc{Jetset} PS~\cite{jetset}, \textsc{Herwig}  
\cite{herwig} and \textsc{Ariadne} \cite{ariadne}) have been used 
for these comparisons. We also compare our measurements with predictions from
QCD models with no coherence effects (\textsc{Cojets} \cite{cojets}).
These Monte Carlo programs use different approaches to describe both the 
perturbative parton shower evolution and non-perturbative hadronisation 
processes. They have been tuned to reproduce the global event shape 
distributions and the charged particle multiplicity distribution measured at 
91.2 \GeV\ \cite{l3tune}. 

The moments of event shape variables are measured between
30 \GeV\ and 189 \GeV. The perturbative and non-perturbative QCD
contributions are obtained from a fit using the power correction 
formula~\cite{qcdpower}. This approach was first applied by the DELPHI
 collaboration \cite{delphi-power}.

The strong coupling constant $\as$ is also determined at each of these 
centre-of-mass energies by comparing the measured event shape 
distributions with predictions of second order QCD calculations 
\cite{qcdz} containing resummed leading and next-to-leading order terms 
\cite{qcdresum}. 

Section \ref{sec:sel} describes the selection of hadronic events. 
Measurements of event shape variables and estimation of systematic 
errors are described in section \ref{sec:evshap}. 
Section \ref{sec:als} presents a comparison of the data with 
predictions from various QCD models, a study of the power correction
ansatz and a determination of $\as$ from event shape distributions.
The results are summarised in section~\ref{sec:summ}.

\section{Event Selection}\label{sec:sel}

The selection of $\ee$ $\rightarrow$ hadrons events is based on the 
energy measured in the electromagnetic and hadron calorimeters.
We use energy clusters in the calorimeters with a 
minimum energy of 100 \MeV. We measure the total visible energy ($\evis$)
and the energy imbalances parallel ($\epar$) and perpendicular 
($\eperp$) to the beam direction. Backgrounds are different 
for hadronic Z decays, hadronic events at reduced centre-of-mass 
energies and at high energies. This is reflected in the different selection 
cuts used for these three types of data sets.

We use Monte Carlo events to estimate the efficiency of the selection 
criteria and purity of the data sample. Monte Carlo events for the process 
$\ee \rightarrow \qq (\gamma )$ have been generated by the parton shower 
programs \textsc{Jetset} and \textsc{Pythia} \cite{pythia} and passed 
through the L3 detector simulation~\cite{l3-simul}. The background 
events are simulated with appropriate event generators: \textsc{Pythia} 
and \textsc{Phojet} \cite{phojet} for two-photon events, \textsc{Koralz}
\cite{koralz} for the $\tau^{+}\tau^{-}(\gamma)$ final state, 
\textsc{Bhagene} \cite{bhagene} and \textsc{Bhwide} \cite{bhwide} for Bhabha 
events, \textsc{Koralw} \cite{koralw} for W-pair production and 
\textsc{Pythia} for Z-pair production.

Details of event selection at $\rs$ $\approx$ $\MZ$ and at reduced 
centre-of-mass energies have been described earlier \cite{l3qcd91,l3qqg}.
At $\rs$ $\approx$ $\MZ$, we have used only a small subset of the complete 
data sample (8.3 $\pb$ out of 142.4 $\pb$ of integrated luminosity) which
still provides an experimental error three times smaller than theoretical
uncertainties.

Data at $\rs$ = 130 and 136 \GeV\  were collected in two separate runs 
during  1995  \cite{l3qcd133} and 1997. 
The main background at these energies comes from ISR resulting in a mass 
of the hadronic system close to $\MZ$. 
This background is reduced by applying a cut in 
the two dimensional plane of $\mid\epar\mid /\evis$ and $\evis /\rs$. In 
the current analysis, data sets from the two years have been combined and 
the cuts are optimised to get the best efficiency times purity.

For the data at $\rs$ $\geq$ 161 \GeV, additional backgrounds arise 
from W-pair and Z-pair production. A substantial fraction ($\sim$ 80\%) of 
these events can be removed by a specific selection \cite{l3qcdlep2}
based on:
\begin{itemize}
\item  forcing the event to a 4-jet topology using the Durham algorithm
             \cite{kt},
\item  performing a kinematic fit imposing the constraints of 
             energy-momentum conservation,
\item  making cuts on energies of the most and the least energetic jets
             and on ${y_{34}^{\mathrm{D}}}$, where ${y_{34}^{\mathrm{D}}}$ is 
             the jet resolution parameter for which the event is classified as 
             a three-jet rather than a four-jet event.
\end{itemize}
These cuts have also been optimised at each energy point.
For centre-of-mass energies at or above 130 \GeV,
hadronic events with ISR photon energy larger than 0.18$\rs$ are
considered as background.

The integrated luminosity, selection efficiency, purity and number of selected 
events for each of the energy points are summarised in Table \ref{tab:events}.

\section{Measurement of Event Shape Variables}\label{sec:evshap}

We measure five global event shape variables for which improved analytical QCD 
calculations \cite{qcdresum} are available.
These are thrust ($T$), scaled heavy jet mass ($\rho$), total ($\bt$) and 
wide ($\bw$) jet broadening variables and the $C$-parameter.

For Monte Carlo events, the global event shape variables are calculated
before (particle level) and after (detector level) detector simulation. The
calculation before detector simulation takes into account all stable
charged and neutral particles.           
The measured distributions at detector level differ from the ones at 
particle level because of detector effects, limited acceptance and
resolution.
After subtracting the background obtained from simulations,
the measured distributions for all energies except $\rs$ $\approx$ $\MZ$
are corrected for detector effects, acceptance and resolution on a 
bin-by-bin basis by comparing the detector level results with the 
particle level results. 
The level of migration is kept at a negligible level with
a bin size larger than the experimental resolution.
At $\rs$ $\approx$ $\MZ$, the detector effects are unfolded for these 
event shape variables using a regularised unfolding method \cite{run}.
We also correct the data for initial and final state photon radiation
bin-by-bin using Monte Carlo distributions at particle level with and 
without radiation. 

The systematic uncertainties in the distributions of event shape variables 
arise mainly due to uncertainties in the estimation of detector correction and
background estimation.
The uncertainty in the detector correction has been estimated by several
independent checks:
\begin{itemize}
 \item The definition of reconstructed objects used to calculate the 
       observables has been changed. Instead of using only calorimetric 
       clusters, the analysis has been repeated with objects obtained from a 
       non-linear combination of energies of charged tracks and calorimetric 
       clusters. At $\rs$ $\approx$ $\MZ$, we use a track based selection and 
       the event shape variables are constructed from the tracks.
 \item The effect of different particle densities in correcting the measured
       distribution has been estimated by changing the signal Monte Carlo
       program (\textsc{Herwig} instead of \textsc{Jetset}).
 \item The acceptance has been reduced by restricting the events to the 
       central part of the detector ($|\cos(\theta_\mathrm{T})|< 0.7$, where 
       $\theta_\mathrm{T}$ is the polar angle of the thrust axis relative 
       to the beam direction) where the energy resolution is better.
\end{itemize}

The uncertainty on the background composition of the selected event sample
has been estimated differently for the three types of data sets. At $\rs$
$\approx$ $\MZ$, the background contamination is negligible and the
uncertainty due to that has been neglected. For data samples at reduced
centre-of-mass energies, the systematic errors arising from background 
subtraction have been estimated \cite{l3qqg} by:
\begin{itemize}
 \item varying, by one standard deviation, the background scale factor which
       takes into account the lack of isolated $\pi^{0}$ and $\eta$ 
       production in the Monte Carlo sample,
 \item varying the cuts on neural network probability,
       jet and local isolation angles, and energy in the local isolation
       cone. 
\end{itemize}
At high energies, the uncertainty is determined by repeating the analysis 
with:
\begin{itemize}
 \item an alternative criterion to reject the hard initial state photon 
       events based on a cut on the kinematically reconstructed effective 
       centre-of-mass energy,
 \item a variation of the estimated two-photon interaction background by 
       $\pm$ 30\% and by changing the background Monte Carlo program
       (\textsc{Phojet} instead of \textsc{Pythia}), and
 \item a variation of the $\WW$ background estimate by changing the W-pair
       rejection criteria.
\end{itemize}

The systematic uncertainties obtained from different sources are combined in
quadrature. At high energies, uncertainties due to ISR and $\WW$ backgrounds
are the most important ones. They are roughly equal and are 2-3 times
larger than the uncertainties due to the detector correction.

Apart from the data set at $\rs$ $\approx$ $\MZ$, statistical fluctuations 
are not negligible in the estimation of systematic effects.
The statistical component of the systematic uncertainty is determined by 
splitting the overall Monte Carlo sample into luminosity weighted 
sub-samples and treating each of these sub-samples as data. The 
spread in the mean position gives an estimate of the statistical
component and is taken out from the original estimate in quadrature.

\section{Results}\label{sec:als}

\subsection{Comparison with QCD models}

\begin{figure}[ht]
\begin{center}
 \includegraphics[width=16.0cm]{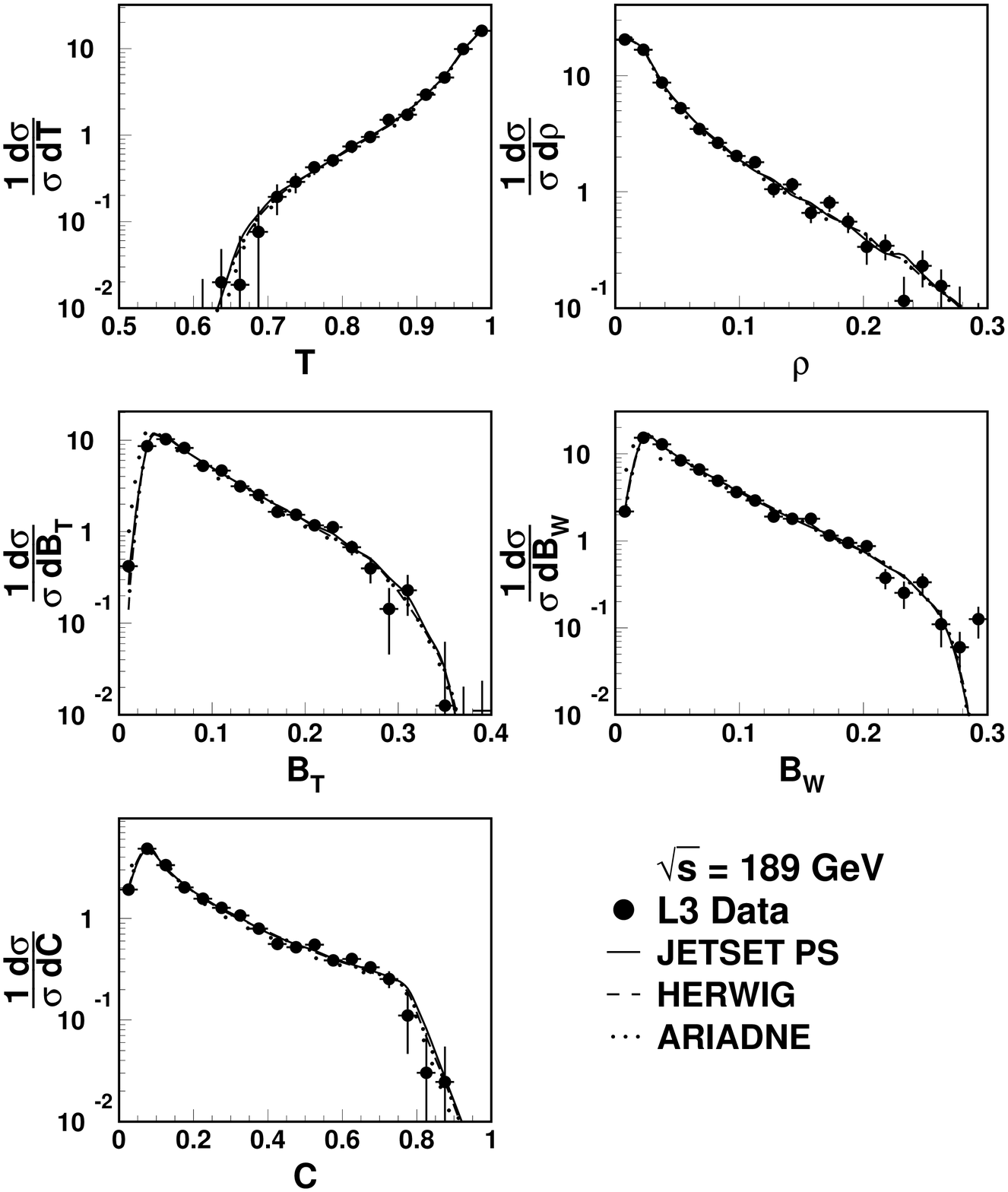}
\end{center}
\caption[]{Distributions for thrust, $T$, scaled heavy jet mass,
           $\rho$, total and wide jet broadenings, $\bt$ and $\bw$, and the 
           $C$-parameter at $\rs$ = 189 \GeV\  in comparison with QCD 
           model predictions. The errors shown are statistical only.}
\label{fig:part}
\end{figure}

Figure~\ref{fig:part}  shows the corrected distributions for thrust,
scaled heavy jet mass, total and wide jet broadening and the $C$-parameter
obtained at $\rs$ = 189 \GeV.
The data are compared with predictions from QCD models \textsc{Jetset} PS, 
\textsc{Herwig} and \textsc{Ariadne} at particle level.
The agreement is satisfactory. 
 
\begin{figure}[htbp]
\begin{center}
    \includegraphics*[width=16.0cm]{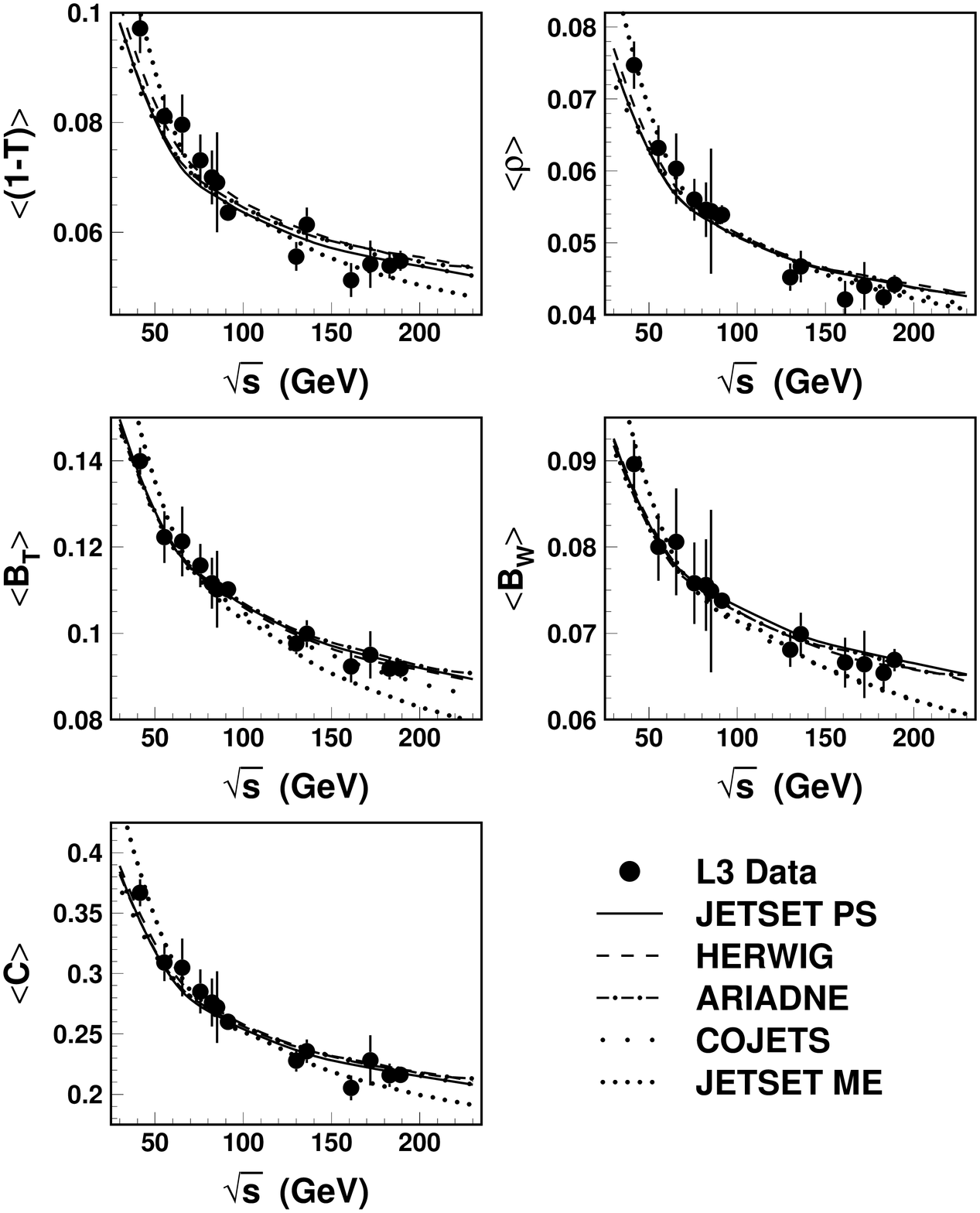}
\end{center}
\caption[]{The first moments of the five event shape 
           variables, $1-T$, $\rho$, $\bt$, $\bw$ and $C$, as a 
           function of the centre-of-mass energy, compared with several 
           QCD models.}
\label{fig:evol}
\end{figure}

An important test of QCD models is a comparison of the energy
evolution of the event shape variables.
The energy dependence  of the mean event shape variables arises mainly
from two sources: the logarithmic energy scale dependence of $\as$ and the
power law behaviour of non-perturbative effects.
The first moments of the five event shape variables are shown in 
Figure~\ref{fig:evol} and Table \ref{tab:fmom}.
Also shown are the energy dependences of these quantities as predicted by
\textsc{Jetset} PS, \textsc{Herwig}, \textsc{Ariadne}, 
\textsc{Cojets} and \textsc{Jetset} ME ($\cal{O}$($\as^{2}$) matrix element
implementation).
All the models with the possible exception of \textsc{Jetset} ME give a
good description of the data.

\subsection{Power Law Correction Analysis}

The energy dependence of moments of the event shape variables has been
described \cite{qcdpower} as a sum of the perturbative contributions and a 
power law dependence due to non-perturbative contributions. The first moment 
of an event shape variable $f$ is written as
\begin{eqnarray}
 \langle f\rangle & = & \langle f_{\mathrm{pert}} \rangle \; +\;
                        \langle f_{\mathrm{pow}}\rangle \; ,
\end{eqnarray}
where the perturbative contribution $\langle f_{\mathrm{pert}}\rangle$ has
been determined to $\cal{O}$$(\as^{2})$ \cite{qcd-event2}.
The power correction term \cite{qcdpower}, for $1-T$, $\rho$, and $C$,
is given by
\begin{eqnarray}
\langle f_{\mathrm{pow}}\rangle & = & c_{f} {\cal{P}} \; ,
\end{eqnarray}
where the factor $c_{f}$ depends on the shape variable $f$ and
${\cal{P}}$ is supposed to have a universal form:
\begin{eqnarray}
 {\cal{P}} & = & {{4C_{F}}\over{\pi^{2}}} {\cal{M}} {{\mu_{I}}\over{\rs}}
\left[ \alpha_{0}(\mu_{I}) - \as(\rs) - \beta_{0}{{\as^{2}(\rs)}\over{2\pi}}
 \left( \ln {{\rs}\over{\mu_{I}}} + {{K}\over{\beta_{0}}} + 1\right)\right]
\end{eqnarray}
for a renormalisation scale fixed at $\rs$.
The parameter $\alpha_{0}$ is related to the value of $\as$ in the 
non-perturbative region below an infrared matching scale $\mu_{I}$
(= 2 \GeV); $\beta_{0}$ is $(11\Nc - 2\Nf)/3$, where $\Nc$ is the number of 
colours and $\Nf$ is the number of active flavours.
$K$ = (67/18 $-$ $\pi^{2}$/6)$C_{A}$ $-$ $5\Nf/9$ and $C_{F}$, $C_{A}$ are 
the usual colour factors.
The Milan factor ${\cal{M}}$ is 1.49 for $\Nf$ = 3. 
For the jet broadening variables, the power correction term takes the form
\begin{eqnarray}
\langle f_{\mathrm{pow}}\rangle & = & c_{f} F {\cal{P}} \; ,
\end{eqnarray}
where
\begin{eqnarray}
 F & = & \left( {{\pi}\over{2\sqrt{a C_{F}\alpha_{\mathrm{CMW}}}}} + 
         {{3}\over{4}} - {{\beta_{0}}\over{6a\ C_{F}}} - 0.6137 +
         {\cal{O}}(\sqrt{\as}) \right)
\end{eqnarray}
and $a$ takes a value 1 for $\bt$ and 2 for $\bw$
and $\alpha_{\mathrm{CMW}}$ is related to $\as$ \cite{qcdpower}.

\begin{figure}[htbp]
\begin{center}
    \includegraphics*[width=16.0cm]{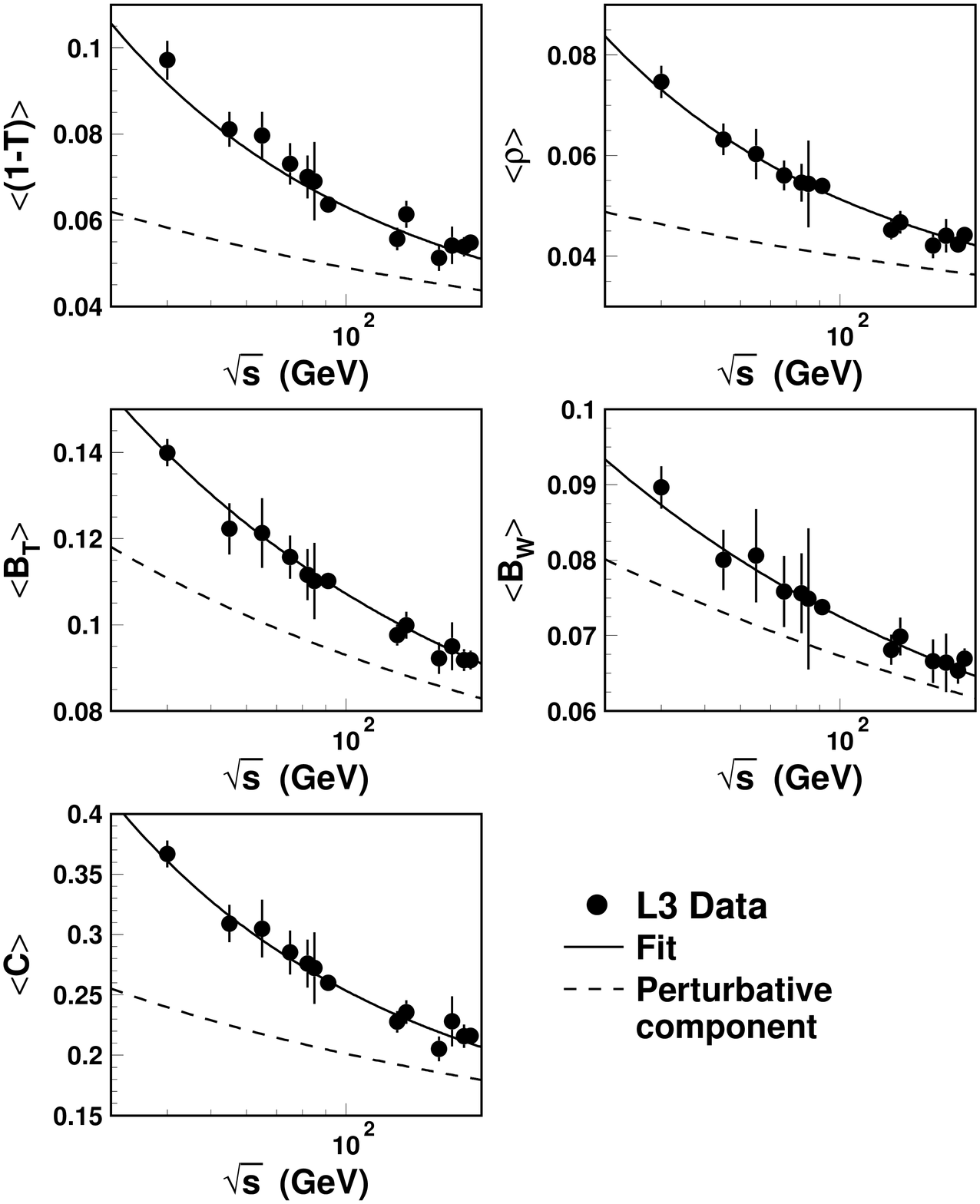}
\end{center}
\caption[]{The first moments of the five event shape variables, $1-T$, $\rho$,
          $\bt$, $\bw$ and $C$ compared to the results of a fit including
          perturbative and power law contributions.}
\label{fig:fmom}
\end{figure}

We have carried out fits to the first moments of the five event shape 
variables separately with $\as (\MZ)$ and $\alpha_{0}$ as free
parameters. 
The diagonal terms of the covariance matrix between the different energy
points are constructed by summing in quadrature the systematic 
uncertainty and the statistical error. 
The off-diagonal terms are obtained from the common systematic errors. 
The results of the fits are summarised in Table \ref{tab:moments} and 
shown in Figure \ref{fig:fmom}. 

The five values of $\alpha_{0}$ obtained from the event shape variables
agree within errors, supporting the predicted universality
of the power law behaviour. 
The theoretical predictions for event shape variables, being
incomplete, give different estimates of $\alpha_{0}$ and $\as$. Since the
measurements are fully correlated, the best estimates of the overall 
values are obtained by taking an unweighted average:
\begin{eqnarray} 
 \alpha_{0} & = & 0.537\  \pm\ 0.070\  \pm\ 0.021 \; ,\\
 \as (\MZ ) & = & 0.1110\ \pm\ 0.0045\ \pm\ 0.0034 \; .
\end{eqnarray}
The first error on each measurement is experimental and is obtained from the 
average of the five errors on $\alpha_{0}$ and $\as$. 
To estimate theoretical uncertainties we vary the renormalisation scale 
between $0.5\rs$ and $2.0\rs$ and $\alpha_{0}$ and $\as (\MZ )$ vary on 
average by $\pm 0.021$ and $\pm 0.0033$ respectively.
A variation of $\mu_{I}$ in the range from 1 to 3 \GeV\ gives an additional 
uncertainty on $\as (\MZ)$ of $\pm$ 0.0010. 
These two estimates of theoretical uncertainties are combined in 
quadrature and quoted as the second error.

\begin{figure}[htbp]
\begin{center}
    \includegraphics*[width=16.0cm]{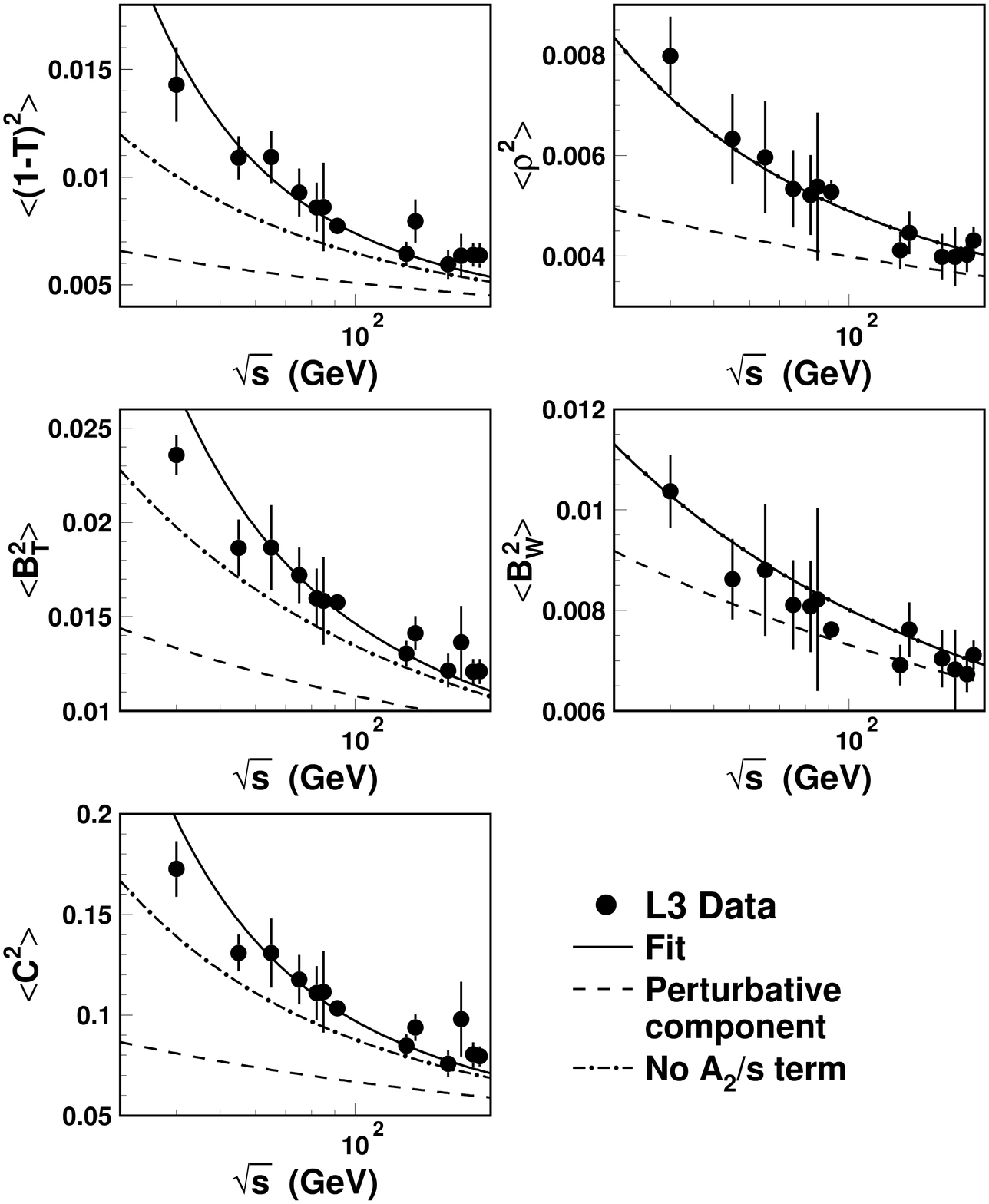}
\end{center}
\caption[]{The second moments of the five event shape variables, $1-T$, $\rho$,
          $\bt$, $\bw$ and $C$ compared to the results of a fit including
          perturbative and power law contributions. The parameters $\alpha_{0}$
          and $\as$ are fixed to the values obtained by the fits to the
          first moments. The $A_{2}/s$ term is negligibly small for $\rho$
          and $\bw$ but is necessary to reproduce the behaviour of $1-T$, 
          $\bt$ and $C$.}
\label{fig:smom}
\end{figure}

We have also measured the second moments of these shape variables which
are summarised in Table \ref{tab:smom}. The energy dependence of these moments 
has been analysed in terms of power law corrections. For variables $1-T$, 
$\rho$ and $C$, the following result is expected to hold \cite{salam}:
\begin{eqnarray}
 \langle f^{2}\rangle & = & \langle f^{2}_{\mathrm{pert}} \rangle \; +\;
            2 \langle f_{\mathrm{pert}}\rangle c_{f}  {\cal{P}} \; +\;
           {\cal{O}}\left( {{1}\over{s}} \right) \; .
\end{eqnarray}
This assumes that the non-perturbative correction to the distributions causes
only a shift. For jet broadenings the power corrections are more complicated. 
The $\cal{O}$$({{1}\over{s}})$ term has been parametrised as $A_{2}/s$
and is expected to be small for $1-T$, $\rho$ and $C$. Fits have been performed
to the second moments where $\alpha_{0}$ and $\as$ have been fixed to the 
values obtained from the corresponding fits to the first moments.
Figure \ref{fig:smom} shows the second moments compared to these fits.
The contributions of the $\cal{O}$$({{1}\over{s}})$ term are 
non-negligible for $1-T$ and $C$, in contradiction with the expectation.
The five values of $A_{2}$, as obtained from the fits, are summarised in
Table~\ref{tab:moments}.

\subsection{{\boldmath $\as$} from Event Shape Distributions}

In order to derive $\as$ from event shape variables at each energy point
we fit the measured distributions 
to theoretical calculations based on $\cal{O}$$(\as^{2})$ perturbative
QCD with resummed leading and next-to-leading order terms.
These calculations are performed at parton level and do not include heavy 
quark mass effects. 
To compare the analytical calculations with the experimental distributions,
the effects of hadronisation and decays have been corrected for using Monte 
Carlo programs.

The fit ranges used take into account the limited 
statistics at high energy as well as the reliability of the resummation 
calculation and are given in Table \ref{tab:als}.
In this analysis, we determine $\as$ at $\rs$ = 130, 136 and 
189 \GeV\  for the first time. 
We also include the measurements done at $\rs$ = 161, 172 and 
183 \GeV\  since the experimental systematic uncertainties are considerably 
reduced by using an improved selection method and by subtracting the 
statistical component of the systematic uncertainties. 
All the measurements are summarised in Table \ref{tab:als}.
These measurements supersede those published previously\cite{l3qcdlep2}.

The experimental errors include the statistical errors and
the experimental systematic uncertainties.
The theoretical error is obtained from estimates \cite{l3qcdlep2} of the 
hadronisation uncertainty and of the errors coming from the uncalculated 
higher orders in the QCD predictions.
The estimate of the theoretical error does not always
reflect the true size of uncalculated higher order terms. An independent
estimate is obtained from a comparison of $\as$ measurements from many event 
shape variables which are
affected differently by higher order corrections and hadronisation effects.
To obtain a combined value for the strong coupling constant we take the 
unweighted average of the five $\as$ values.
We estimate the overall theoretical error from the simple average of the
five theoretical errors or from half of the maximum spread in the five
$\as$ values. Both estimates yield similar results. The combined
results are summarised in Table \ref{tab:alscomb}. The earlier measurements
at $\rs$ = $\MZ$ and at reduced centre-of-mass energies determined $\as$
from four event shape variables only: $T$, $\rho$, $\bt$ and $\bw$. For 
comparison we also provide in Table \ref{tab:alscomb} the mean from these 
four measurements.

\begin{figure}[htbp]
\begin{center}
    \includegraphics[width=14cm]{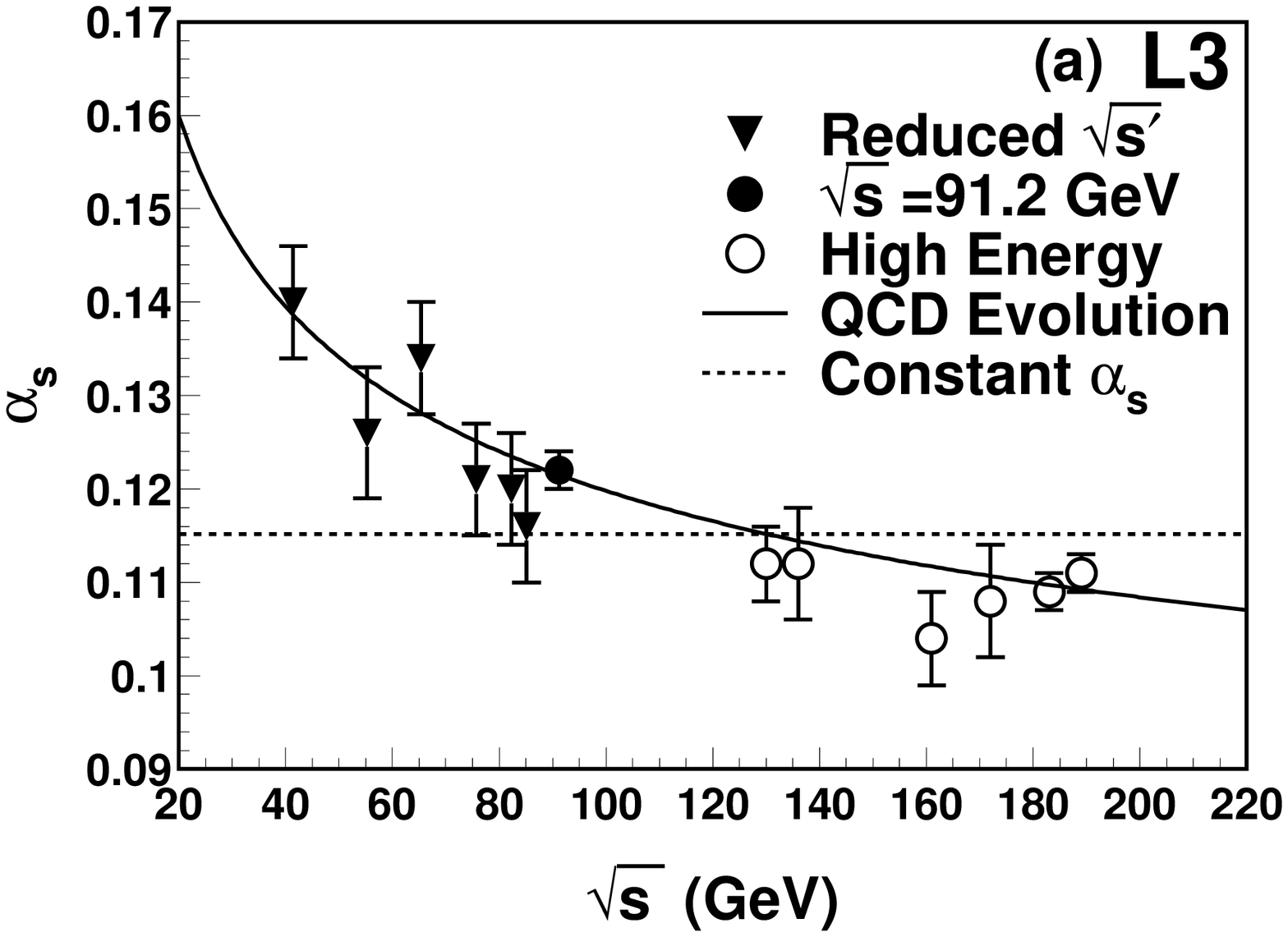}
    \includegraphics[width=14cm]{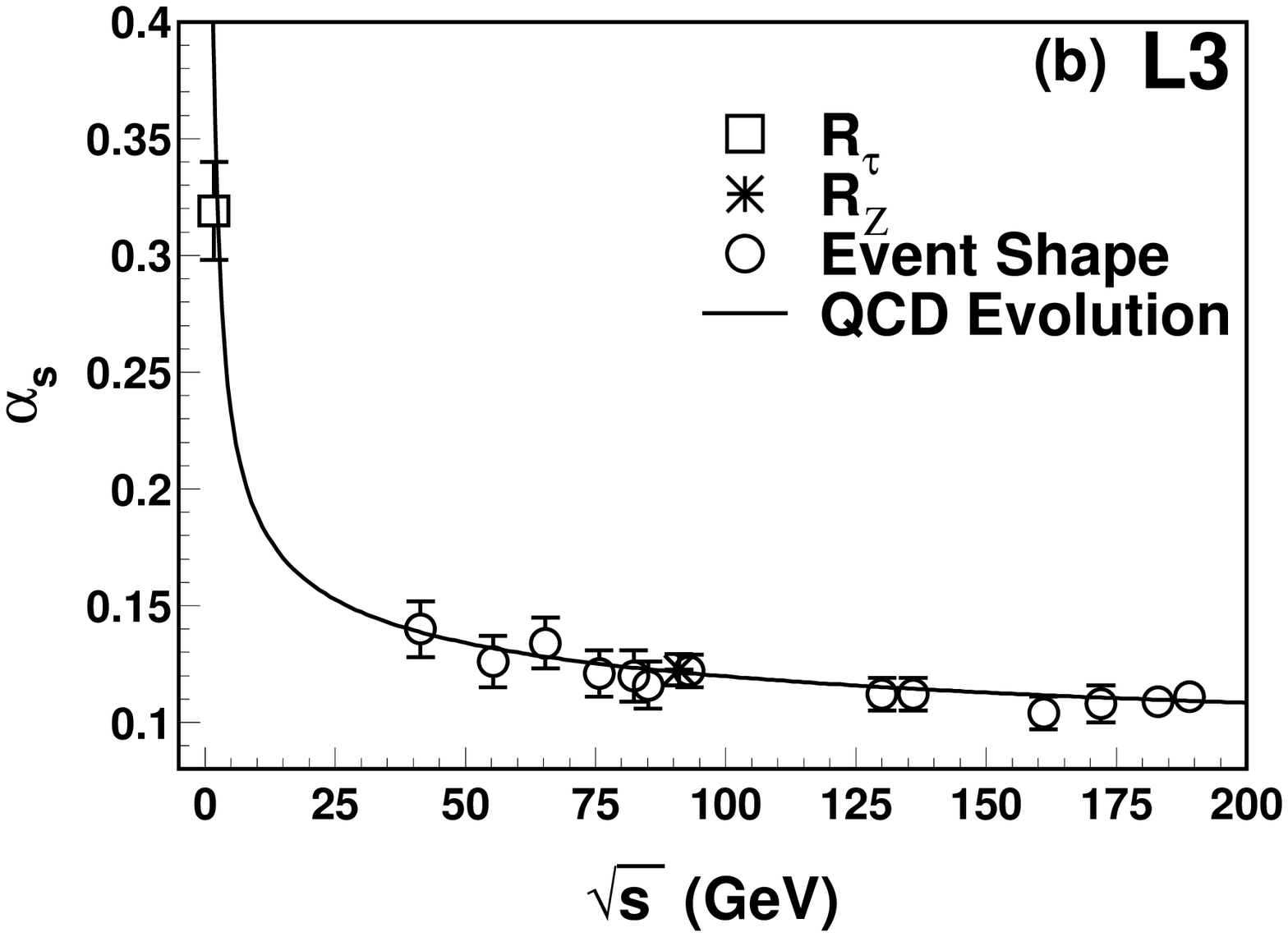}
\end{center}
\caption[]{a) $\as$  measurements from event shape distributions as a function
          of the centre-of-mass energy. The errors shown are experimental
          only. The solid and dashed lines are fits with the energy dependence 
          of $\as$ as expected from QCD and with constant $\as$, respectively.
          \\ b) $\as$ values as determined by L3 from the $\tau$ lifetime
          measurement, Z lineshape and event shape distributions. The line is a
          fit to the QCD evolution function to the measurements made from
          event shape variables.}
\label{fig:alsevol}
\end{figure}

We compare the energy dependence of the measured $\as$ values with the
prediction from QCD in Figure \ref{fig:alsevol}a.
The theoretical errors are strongly correlated 
between these  measurements. 
The error appropriate to a measurement of the energy dependence of 
$\as$ can then be considered to be experimental.
The experimental systematic errors on $\as$ are dominated by the background
uncertainties. These are similar for all the individual low energy or high
energy data points but differ between the low energy, Z peak and high energy
data sets. The experimental systematic errors are then different and 
uncorrelated between the three data sets, but are taken as fully correlated
between individual low energy or high energy measurements.
The thirteen measurements in Figure~\ref{fig:alsevol}a are shown with 
experimental errors only, together with a fit to the QCD evolution equation
\cite{pdg} with $\as(\MZ)$ as a free parameter.
The fit gives a $\chi^2$ of 13.5 for 12 degrees of
freedom corresponding to a confidence level of 0.34 with a fitted value
of $\as$:
\begin{eqnarray}
  \as (\MZ) & = & 0.1215~\pm~0.0012~\pm~0.0061 \; .
\end{eqnarray}
The first error is experimental and the second error is theoretical. 
On the other hand, a fit with constant $\as$ gives a $\chi^2$ of 65.1
for 12 degrees of freedom.
The value of $\as (\MZ)$ thus obtained is in agreement with the value
obtained in the power law ansatz analysis considering the experimental 
and the theoretical uncertainties.

Figure~\ref{fig:alsevol}b summarises the $\as$ values determined by L3  
from the $\tau$ lifetime measurement~\cite{l3tau}, Z 
lineshape~\cite{l3lineshape}
and event shape distributions at various energies, together with the QCD 
prediction obtained from a fit to the event shape measurements only.
These measurements support the energy evolution of the strong
coupling constant predicted by QCD.

The slope in the energy evolution of $\as$ depends on the number of active
flavours. We have performed a fit with $\Nf$ as a free
parameter along with $\as$ and obtain the number of active flavours:
\begin{eqnarray}
  \Nf & = & 5.0 ~\pm~ 1.3 ~\pm~ 2.0 \; ,
\end{eqnarray}
where the first error is experimental and the second is due to theoretical
uncertainties. The errors have been estimated by using the covariance matrix
determined from experimental and overall errors on $\as$ in the fit.
This result agrees with the expectation $\Nf$ = 5.

\section{Summary}\label{sec:summ}

We have measured distributions of event shape variables in hadronic
events from $\ee$ annihilation at centre-of-mass energies from 30 \GeV\ to 189 
\GeV. These distributions as well as the energy 
dependence of their first moments are well described by parton shower models.

The energy dependence of the first two moments has been compared with 
second order perturbative QCD with power law corrections for the 
non-perturbative effects. The fits of the five event shape variables agree 
with a universal power law behaviour giving $\alpha_{0}$ = 0.537 $\pm$ 0.070
(exp) $\pm$ 0.021 (th). We find a non-negligible contribution from an
$\cal{O}$$({{1}\over{s}})$ term  in describing the second moments
of $1-T$, $\bt$  and $C$.

The event shape distributions are compared to second order QCD calculations
together with resummed leading and next-to-leading log terms.  The data are 
well described by these calculations at all energies. The measurements 
demonstrate the running of $\as$ as expected in QCD with a value of  
$\as (\MZ)$ = 0.1215 $\pm$ 0.0012 (exp) $\pm$ 0.0061 (th).
From the energy dependence of $\as$, we determine the number of active 
flavours to be  $\Nf$ =  5.0 $\pm$ 1.3 (exp) $\pm$~2.0~(th).

\section{Acknowledgments}

We express our gratitude to the CERN accelerator divisions for the excellent
performance of the LEP machine. We acknowledge with appreciation the effort
of the engineers, technicians and support staff who have participated in the
construction and maintenance of this experiment.

%
%
\newpage
\section*{Author List}
\typeout{   }     
\typeout{Using author list for paper 208 -- ? }
\typeout{$Modified: Tue May  2 13:45:26 2000 by clare $}
\typeout{!!!!  This should only be used with document option a4p!!!!}
\typeout{   }
%
%
%
%
%
%

\newcount\tutecount  \tutecount=0
\def\tutenum#1{\global\advance\tutecount by 1 \xdef#1{\the\tutecount}}
\def\tute#1{$^{#1}$}
\tutenum\aachen            
\tutenum\nikhef            
\tutenum\mich              
\tutenum\lapp              
\tutenum\basel             
\tutenum\lsu               
\tutenum\beijing           
\tutenum\berlin            
\tutenum\bologna           
\tutenum\tata              
\tutenum\ne                
\tutenum\bucharest         
\tutenum\budapest          
\tutenum\mit               
\tutenum\debrecen          
\tutenum\florence          
\tutenum\cern              
\tutenum\wl                
\tutenum\geneva            
\tutenum\hefei             
\tutenum\seft              
\tutenum\lausanne          
\tutenum\lecce             
\tutenum\lyon              
\tutenum\madrid            
\tutenum\milan             
\tutenum\moscow            
\tutenum\naples            
\tutenum\cyprus            
\tutenum\nymegen           
\tutenum\caltech           
\tutenum\perugia           
\tutenum\cmu               
\tutenum\prince            
\tutenum\rome              
\tutenum\peters            
\tutenum\potenza           
\tutenum\salerno           
\tutenum\ucsd              
\tutenum\santiago          
\tutenum\sofia             
\tutenum\korea             
\tutenum\alabama           
\tutenum\utrecht           
\tutenum\purdue            
\tutenum\psinst            
\tutenum\zeuthen           
\tutenum\eth               
\tutenum\hamburg           
\tutenum\taiwan            
\tutenum\tsinghua          

{
\parskip=0pt
\noindent
{\bf The L3 Collaboration:}
\ifx\selectfont\undefined
 \baselineskip=10.8pt
 \baselineskip\baselinestretch\baselineskip
 \normalbaselineskip\baselineskip
 \ixpt
\else
 \fontsize{9}{10.8pt}\selectfont
\fi
\medskip
\tolerance=10000
\hbadness=5000
\raggedright
\hsize=162truemm\hoffset=0mm
\def\r{\rlap,}
\noindent

M.Acciarri\r\tute\milan\
P.Achard\r\tute\geneva\ 
O.Adriani\r\tute{\florence}\ 
M.Aguilar-Benitez\r\tute\madrid\ 
J.Alcaraz\r\tute\madrid\ 
G.Alemanni\r\tute\lausanne\
J.Allaby\r\tute\cern\
A.Aloisio\r\tute\naples\ 
M.G.Alviggi\r\tute\naples\
G.Ambrosi\r\tute\geneva\
% ...
\newpage
\begin{list}{A}{\itemsep=0pt plus 0pt minus 0pt\parsep=0pt plus 0pt minus 0pt
                \topsep=0pt plus 0pt minus 0pt}
\item[\aachen]
 I. Physikalisches Institut, RWTH, D-52056 Aachen, FRG$^{\S}$\\
 III. Physikalisches Institut, RWTH, D-52056 Aachen, FRG$^{\S}$
\item[\nikhef] National Institute for High Energy Physics, NIKHEF, 
     and University of Amsterdam, NL-1009 DB Amsterdam, The Netherlands
\item[\mich] University of Michigan, Ann Arbor, MI 48109, USA
\item[\lapp] Laboratoire d'Annecy-le-Vieux de Physique des Particules, 
     LAPP,IN2P3-CNRS, BP 110, F-74941 Annecy-le-Vieux CEDEX, France
\item[\basel] Institute of Physics, University of Basel, CH-4056 Basel,
     Switzerland
\item[\lsu] Louisiana State University, Baton Rouge, LA 70803, USA
\item[\beijing] Institute of High Energy Physics, IHEP, 
  100039 Beijing, China$^{\triangle}$ 
\item[\berlin] Humboldt University, D-10099 Berlin, FRG$^{\S}$
\item[\bologna] University of Bologna and INFN-Sezione di Bologna, 
     I-40126 Bologna, Italy
\item[\tata] Tata Institute of Fundamental Research, Bombay 400 005, India
\item[\ne] Northeastern University, Boston, MA 02115, USA
\item[\bucharest] Institute of Atomic Physics and University of Bucharest,
     R-76900 Bucharest, Romania
\item[\budapest] Central Research Institute for Physics of the 
     Hungarian Academy of Sciences, H-1525 Budapest 114, Hungary$^{\ddag}$
\item[\mit] Massachusetts Institute of Technology, Cambridge, MA 02139, USA
\item[\debrecen] KLTE-ATOMKI, H-4010 Debrecen, Hungary$^\P$
\item[\florence] INFN Sezione di Firenze and University of Florence, 
     I-50125 Florence, Italy
\item[\cern] European Laboratory for Particle Physics, CERN, 
     CH-1211 Geneva 23, Switzerland
\item[\wl] World Laboratory, FBLJA  Project, CH-1211 Geneva 23, Switzerland
\item[\geneva] University of Geneva, CH-1211 Geneva 4, Switzerland
\item[\hefei] Chinese University of Science and Technology, USTC,
      Hefei, Anhui 230 029, China$^{\triangle}$
\item[\seft] SEFT, Research Institute for High Energy Physics, P.O. Box 9,
      SF-00014 Helsinki, Finland
\item[\lausanne] University of Lausanne, CH-1015 Lausanne, Switzerland
\item[\lecce] INFN-Sezione di Lecce and Universit\'a Degli Studi di Lecce,
     I-73100 Lecce, Italy
\item[\lyon] Institut de Physique Nucl\'eaire de Lyon, 
     IN2P3-CNRS,Universit\'e Claude Bernard, 
     F-69622 Villeurbanne, France
\item[\madrid] Centro de Investigaciones Energ{\'e}ticas, 
     Medioambientales y Tecnolog{\'\i}cas, CIEMAT, E-28040 Madrid,
     Spain${\flat}$ 
\item[\milan] INFN-Sezione di Milano, I-20133 Milan, Italy
\item[\moscow] Institute of Theoretical and Experimental Physics, ITEP, 
     Moscow, Russia
\item[\naples] INFN-Sezione di Napoli and University of Naples, 
     I-80125 Naples, Italy
\item[\cyprus] Department of Natural Sciences, University of Cyprus,
     Nicosia, Cyprus
\item[\nymegen] University of Nijmegen and NIKHEF, 
     NL-6525 ED Nijmegen, The Netherlands
\item[\caltech] California Institute of Technology, Pasadena, CA 91125, USA
\item[\perugia] INFN-Sezione di Perugia and Universit\'a Degli 
     Studi di Perugia, I-06100 Perugia, Italy   
\item[\cmu] Carnegie Mellon University, Pittsburgh, PA 15213, USA
\item[\prince] Princeton University, Princeton, NJ 08544, USA
\item[\rome] INFN-Sezione di Roma and University of Rome, ``La Sapienza",
     I-00185 Rome, Italy
\item[\peters] Nuclear Physics Institute, St. Petersburg, Russia
\item[\potenza] INFN-Sezione di Napoli and University of Potenza, 
     I-85100 Potenza, Italy
\item[\salerno] University and INFN, Salerno, I-84100 Salerno, Italy
\item[\ucsd] University of California, San Diego, CA 92093, USA
\item[\santiago] Dept. de Fisica de Particulas Elementales, Univ. de Santiago,
     E-15706 Santiago de Compostela, Spain
\item[\sofia] Bulgarian Academy of Sciences, Central Lab.~of 
     Mechatronics and Instrumentation, BU-1113 Sofia, Bulgaria
\item[\korea]  Laboratory of High Energy Physics, 
     Kyungpook National University, 702-701 Taegu, Republic of Korea
\item[\alabama] University of Alabama, Tuscaloosa, AL 35486, USA
\item[\utrecht] Utrecht University and NIKHEF, NL-3584 CB Utrecht, 
     The Netherlands
\item[\purdue] Purdue University, West Lafayette, IN 47907, USA
\item[\psinst] Paul Scherrer Institut, PSI, CH-5232 Villigen, Switzerland
\item[\zeuthen] DESY, D-15738 Zeuthen, 
     FRG
\item[\eth] Eidgen\"ossische Technische Hochschule, ETH Z\"urich,
     CH-8093 Z\"urich, Switzerland
\item[\hamburg] University of Hamburg, D-22761 Hamburg, FRG
\item[\taiwan] National Central University, Chung-Li, Taiwan, China
\item[\tsinghua] Department of Physics, National Tsing Hua University,
      Taiwan, China
\item[\S]  Supported by the German Bundesministerium 
        f\"ur Bildung, Wissenschaft, Forschung und Technologie
\item[\ddag] Supported by the Hungarian OTKA fund under contract
numbers T019181, F023259 and T024011.
\item[\P] Also supported by the Hungarian OTKA fund under contract
  numbers T22238 and T026178.
\item[$\flat$] Supported also by the Comisi\'on Interministerial de Ciencia y 
        Tecnolog{\'\i}a.
\item[$\sharp$] Also supported by CONICET and Universidad Nacional de La Plata,
        CC 67, 1900 La Plata, Argentina.
\item[$\diamondsuit$] Also supported by Panjab University, Chandigarh-160014, 
        India.
\item[$\triangle$] Supported by the National Natural Science
  Foundation of China.
\end{list}
}
\vfill


\clearpage

\begin{table}[htbp]
\begin{center}\begin{tabular}{|c|c|c|c|c|}\hline
$\rs$   & Integrated & Selection & Sample & Selected       \\
        & Luminosity & Efficiency& Purity & events  \\
(\GeV)  &  ($\pb$)   &   (\%)    &  (\%)  &             \\ \hline
30$-$50 & 142.4      &  48.3     & 68.4   &    1247     \\
50$-$60 & 142.4      &  41.0     & 78.0   &    1047     \\
60$-$70 & 142.4      &  35.2     & 86.0   &    1575     \\
70$-$80 & 142.4      &  29.9     & 89.0   &    2938     \\
80$-$84 & 142.4      &  27.4     & 90.5   &    2091     \\
84$-$86 & 142.4      &  27.5     & 87.0   &    1607     \\
 91.2   &  8.3       &  98.5     & 99.8   &  248100     \\
 130    &   6.1      &  90.0     & 80.6   &     556     \\
 136    &   5.9      &  89.0     & 81.5   &     414     \\
 161    &  10.8      &  89.0     & 81.2   &     424     \\
 172    &  10.2      &  84.8     & 82.6   &     325     \\
 183    &  55.3      &  84.2     & 82.4   &    1500     \\
 189    & 176.8      &  87.8     & 81.1   &    4479     \\
\hline\end{tabular}\end{center}
\caption[]{Summary of integrated luminosity, selection efficiency, sample 
           purity and number of selected hadronic events at the different 
           energies used in this analysis. The energies below $\rs$~=~91 
           \GeV\ are obtained from the full data sample at the Z peak, by
           selecting events with an isolated high energy photon.}
\label{tab:events}
\end{table}

\begin{sidewaystable}[htbp]
\begin{center}
\small{\begin{tabular}{|c|c|c|c|c|c|}\hline
 $\rs$ & \multicolumn{5}{c|}{First moments of} \\ \cline{2-6}
 (\GeV) &  $1-T$   & $\rho$  &  $\bt$    & $\bw$  &  $C$     \\\hline
30$-$50&
  .0971  $\pm$  .0030  $\pm$  .0034 &  
  .0747  $\pm$  .0023  $\pm$  .0023 & 
  .1399  $\pm$  .0027  $\pm$  .0016 &  
  .0896  $\pm$  .0021  $\pm$  .0018 &  
  .3667  $\pm$  .0084  $\pm$  .0073 \\ \hline
50$-$60&
  .0811  $\pm$  .0027  $\pm$  .0029 &  
  .0632  $\pm$  .0021  $\pm$  .0023 & 
  .1223  $\pm$  .0025  $\pm$  .0054 &  
  .0800  $\pm$  .0020  $\pm$  .0034 &  
  .3091  $\pm$  .0080  $\pm$  .0131 \\ \hline
60$-$70&
  .0796  $\pm$  .0021  $\pm$  .0051 &  
  .0603  $\pm$  .0015  $\pm$  .0047 & 
  .1213  $\pm$  .0019  $\pm$  .0079 &  
  .0806  $\pm$  .0014  $\pm$  .0060 &  
  .3049  $\pm$  .0059  $\pm$  .0232 \\ \hline 
70$-$80&
  .0731  $\pm$  .0015  $\pm$  .0045 &  
  .0560  $\pm$  .0011  $\pm$  .0027 & 
  .1157  $\pm$  .0015  $\pm$  .0048 &  
  .0758  $\pm$  .0011  $\pm$  .0046 &  
  .2851  $\pm$  .0044  $\pm$  .0177 \\  \hline
80$-$84&
  .0700  $\pm$  .0018  $\pm$  .0046 &  
  .0546  $\pm$  .0015  $\pm$  .0035 & 
  .1116  $\pm$  .0017  $\pm$  .0057 &  
  .0756  $\pm$  .0014  $\pm$  .0051 &  
  .2759  $\pm$  .0055  $\pm$  .0191 \\  \hline
84$-$86&
  .0691  $\pm$  .0022  $\pm$  .0088 &  
  .0544  $\pm$  .0017  $\pm$  .0085 & 
  .1102  $\pm$  .0021  $\pm$  .0086 &  
  .0749  $\pm$  .0017  $\pm$  .0092 &  
  .2722  $\pm$  .0068  $\pm$  .0289 \\ \hline 
91.2&
  .0636  $\pm$  .0003  $\pm$  .0013 &  
  .0539  $\pm$  .0002  $\pm$  .0013 & 
  .1102  $\pm$  .0002  $\pm$  .0011 &  
  .0738  $\pm$  .0001  $\pm$  .0008 &  
  .2599  $\pm$  .0004  $\pm$  .0054 \\ \hline 
130&
  .0556  $\pm$  .0022  $\pm$  .0014 &  
  .0452  $\pm$  .0018  $\pm$  .0007 & 
  .0976  $\pm$  .0023  $\pm$  .0008 &  
  .0681  $\pm$  .0019  $\pm$  .0007 &  
  .2277  $\pm$  .0072  $\pm$  .0052 \\ \hline 
136&
  .0614  $\pm$  .0029  $\pm$  .0011 &  
  .0467  $\pm$  .0022  $\pm$  .0004 & 
  .0999  $\pm$  .0029  $\pm$  .0011 &  
  .0699  $\pm$  .0024  $\pm$  .0006 &  
  .2357  $\pm$  .0089  $\pm$  .0038 \\ \hline 
161&
  .0513  $\pm$  .0030  $\pm$  .0008 & 
  .0421  $\pm$  .0025  $\pm$  .0007 & 
  .0923  $\pm$  .0032  $\pm$  .0018 &  
  .0666  $\pm$  .0027  $\pm$  .0010 &  
  .2052  $\pm$  .0098  $\pm$  .0028 \\ \hline 
172&
  .0542  $\pm$  .0037  $\pm$  .0022 &  
  .0440  $\pm$  .0028  $\pm$  .0018 & 
  .0950  $\pm$  .0046  $\pm$  .0031 &  
  .0664  $\pm$  .0031  $\pm$  .0023 &  
  .2281  $\pm$  .0159  $\pm$  .0133 \\  \hline
183&
  .0539  $\pm$  .0020  $\pm$  .0011 &  
  .0424  $\pm$  .0014  $\pm$  .0004 & 
  .0918  $\pm$  .0020  $\pm$  .0015 &  
  .0654  $\pm$  .0015  $\pm$  .0010 &  
  .2157  $\pm$  .0063  $\pm$  .0073 \\  \hline
189&
  .0548  $\pm$  .0013  $\pm$  .0013 &  
  .0442  $\pm$  .0009  $\pm$  .0009 & 
  .0918  $\pm$  .0013  $\pm$  .0018 &  
  .0669  $\pm$  .0009  $\pm$  .0010 &  
  .2160  $\pm$  .0040  $\pm$  .0041 \\  \hline
\end{tabular}}
\end{center}
\caption[]{First moments of the five event shape variables at different
           energy points. The two errors are respectively statistical 
           and systematic.}
\label{tab:fmom}
\end{sidewaystable}

\begin{table}[htbp]
\begin{center}
\begin{tabular}{|c|c|c|c|c|} \hline
Observable & $\alpha_{0}$ & $\as (\MZ)$ & $\chi^{2}$/d.o.f.
           & $A_{2}$ (\GeV$^{2}$) \\ \hline
$1-T $ & 0.633 $\pm$ 0.097 & 0.1104 $\pm$ 0.0065 & 11.5/11 & 5.47 $\pm$ 0.56 \\
$\rho$ & 0.523 $\pm$ 0.063 & 0.1027 $\pm$ 0.0050 & 5.5/11  & 0.00 $^{+0.01}_{-0.00}$ \\
$\bt $ & 0.517 $\pm$ 0.044 & 0.1160 $\pm$ 0.0029 & 3.5/11  &13.75 $\pm$ 0.88 \\
$\bw $ & 0.476 $\pm$ 0.100 & 0.1134 $\pm$ 0.0042 & 4.1/11  & 0.00 $^{+0.05}_{-0.00}$ \\
$C$    & 0.537 $\pm$ 0.044 & 0.1125 $\pm$ 0.0038 & 6.3/11  &11.58 $\pm$ 0.88 \\
\hline\end{tabular}\end{center}
\caption[]{Determination of $\alpha_{0}$ and $\as (\MZ)$ from
           fits to the first moments of the event shape distributions
           together with $\chi^{2}$/d.o.f. from those fits. Also shown
           is the $A_{2}$ parameter from fits to the second moments.}
\label{tab:moments}
\end{table}

\begin{sidewaystable}[htbp]
\begin{center}
\small{\begin{tabular}{|c|c|c|c|c|c|}\hline
 $\rs$ & \multicolumn{5}{c|}{Second  moments of} \\ \cline{2-6}
 (\GeV) &  $1-T$   & $\rho$  &  $\bt$    & $\bw$  &  $C$     \\\hline
30$-$50&
     .0143  $\pm$  .0009  $\pm$  .0015 &            
     .0080  $\pm$  .0006  $\pm$  .0005 &            
     .0236  $\pm$  .0009  $\pm$  .0005 &            
     .0104  $\pm$  .0005  $\pm$  .0005 &            
     .1726  $\pm$  .0078  $\pm$  .0115 \\\hline     
50$-$60&
     .0109  $\pm$  .0008  $\pm$  .0006 &            
     .0063  $\pm$  .0005  $\pm$  .0008 &            
     .0187  $\pm$  .0008  $\pm$  .0012 &            
     .0086  $\pm$  .0005  $\pm$  .0006 &            
     .1308  $\pm$  .0066  $\pm$  .0063 \\\hline     
60$-$70&
     .0109  $\pm$  .0006  $\pm$  .0010 &            
     .0060  $\pm$  .0004  $\pm$  .0011 &            
     .0187  $\pm$  .0006  $\pm$  .0022 &            
     .0088  $\pm$  .0003  $\pm$  .0013 &            
     .1308  $\pm$  .0050  $\pm$  .0164 \\\hline     
70$-$80&
     .0093  $\pm$  .0004  $\pm$  .0010 &            
     .0053  $\pm$  .0002  $\pm$  .0007 &            
     .0172  $\pm$  .0005  $\pm$  .0014 &            
     .0081  $\pm$  .0003  $\pm$  .0008 &            
     .1176  $\pm$  .0037  $\pm$  .0117 \\\hline     
80$-$84&
     .0086  $\pm$  .0005  $\pm$  .0010 &            
     .0052  $\pm$  .0003  $\pm$  .0007 &            
     .0160  $\pm$  .0006  $\pm$  .0015 &            
     .0081  $\pm$  .0003  $\pm$  .0008 &            
     .1110  $\pm$  .0047  $\pm$  .0125 \\\hline     
84$-$86&
     .0086  $\pm$  .0006  $\pm$  .0020 &            
     .0054  $\pm$  .0004  $\pm$  .0014 &            
     .0158  $\pm$  .0007  $\pm$  .0022 &            
     .0082  $\pm$  .0004  $\pm$  .0018 &            
     .1115  $\pm$  .0058  $\pm$  .0195 \\\hline     
91.2&
     .0077  $\pm$  .0001  $\pm$  .0003 &            
     .0053  $\pm$  .0001  $\pm$  .0002 &            
     .0158  $\pm$  .0001  $\pm$  .0003 &            
     .0076  $\pm$  .0001  $\pm$  .0002 &            
     .1034  $\pm$  .0003  $\pm$  .0031 \\\hline     
130&
     .0064  $\pm$  .0005  $\pm$  .0002 &            
     .0041  $\pm$  .0003  $\pm$  .0001 &            
     .0131  $\pm$  .0006  $\pm$  .0002 &            
     .0069  $\pm$  .0004  $\pm$  .0001 &            
     .0848  $\pm$  .0050  $\pm$  .0025 \\\hline     
136&
     .0080  $\pm$  .0008  $\pm$  .0007 &            
     .0045  $\pm$  .0004  $\pm$  .0001 &            
     .0141  $\pm$  .0008  $\pm$  .0004 &            
     .0076  $\pm$  .0005  $\pm$  .0002 &            
     .0938  $\pm$  .0064  $\pm$  .0017 \\\hline     
161&
     .0059  $\pm$  .0007  $\pm$  .0002 &            
     .0040  $\pm$  .0004  $\pm$  .0001 &            
     .0121  $\pm$  .0008  $\pm$  .0004 &            
     .0070  $\pm$  .0005  $\pm$  .0002 &            
     .0757  $\pm$  .0064  $\pm$  .0019 \\\hline     
172&
     .0064  $\pm$  .0009  $\pm$  .0005 &            
     .0040  $\pm$  .0005  $\pm$  .0003 &            
     .0136  $\pm$  .0014  $\pm$  .0013 &            
     .0068  $\pm$  .0006  $\pm$  .0005 &            
     .0979  $\pm$  .0133  $\pm$  .0129 \\\hline     
183&
     .0064  $\pm$  .0005  $\pm$  .0001 &            
     .0042  $\pm$  .0003  $\pm$  .0002 &            
     .0121  $\pm$  .0006  $\pm$  .0003 &            
     .0067  $\pm$  .0003  $\pm$  .0002 &            
     .0804  $\pm$  .0051  $\pm$  .0032 \\\hline     
189&
     .0064  $\pm$  .0004  $\pm$  .0004 &         
     .0043  $\pm$  .0002  $\pm$  .0002 &          
     .0121  $\pm$  .0004  $\pm$  .0005 &          
     .0071  $\pm$  .0002  $\pm$  .0002 &          
     .0794  $\pm$  .0032  $\pm$  .0038 \\\hline   
\end{tabular}}
\end{center}
\caption[]{Second moments of the five event shape variables at different
           energy points. The two errors are respectively statistical 
           and systematic.}
\label{tab:smom}
\end{sidewaystable}

\begin{table}[htbp]
\begin{center}
{\small
\begin{tabular}{|l|c|c|c|c|c|}\hline
          & ($1-T$) & $\rho$ & $\bt$ & $\bw$ & $C$ \\ \hline
Fit Range         & 0.00$-$0.30 & 0.00$-$0.20 & 0.00$-$0.25 
                  & 0.00$-$0.20 & 0.05$-$0.50 \\ \hline
\hline
$\as$(130 \GeV)    & 0.1139 & 0.1134 & 0.1153 & 0.1063 & 0.1151 \\\hline
Statistical error & $\pm 0.0036$ & $\pm 0.0034$ & $\pm 0.0027$ 
                  & $\pm 0.0027$ & $\pm 0.0036$ \\
Systematic error  & $\pm 0.0028$ & $\pm 0.0029$ & $\pm 0.0016$ 
                  & $\pm 0.0015$ & $\pm 0.0018$ \\ \hline
Overall experimental error 
                  & $\pm 0.0046$ & $\pm 0.0045$ & $\pm 0.0031$ 
                  & $\pm 0.0031$ & $\pm 0.0040$ \\ \hline
Overall theoretical error 
                  & $\pm 0.0056$ & $\pm 0.0038$ & $\pm 0.0062$ 
                  & $\pm 0.0088$ & $\pm 0.0066$ \\ \hline
$\chi^{2}$/d.o.f. & 6.9 / 10 & 8.4 / 9 & 9.1 / 11 & 12.0 / 12 & 8.5 / 8\\
\hline\hline
$\as$(136 \GeV)    & 0.1166 & 0.1112 & 0.1141 & 0.1045 & 0.1089 \\\hline
Statistical error & $\pm 0.0047$ & $\pm 0.0037$ & $\pm 0.0034$ 
                  & $\pm 0.0032$ & $\pm 0.0043$ \\
Systematic error  & $\pm 0.0024$ & $\pm 0.0013$ & $\pm 0.0010$ 
                  & $\pm 0.0026$ & $\pm 0.0020$ \\ \hline
Overall experimental error 
                  & $\pm 0.0053$ & $\pm 0.0039$ & $\pm 0.0035$ 
                  & $\pm 0.0041$ & $\pm 0.0047$ \\ \hline
Overall theoretical error 
                  & $\pm 0.0060$ & $\pm 0.0037$ & $\pm 0.0064$ 
                  & $\pm 0.0078$ & $\pm 0.0076$ \\ \hline
$\chi^{2}$/d.o.f. & 10.2 / 9 & 11.4 / 13 & 7.7 / 11 & 7.9 / 12 & 11.8 / 8 \\
\hline \hline
$\as$(161 \GeV)    & 0.1018 & 0.1012 & 0.1101 & 0.1032 & 0.1043 \\\hline
Statistical error & $\pm 0.0051$ & $\pm 0.0052$ & $\pm 0.0039$ 
                  & $\pm 0.0039$ & $\pm 0.0055$ \\
Systematic error  & $\pm 0.0022$ & $\pm 0.0022$ & $\pm 0.0015$ 
                  & $\pm 0.0044$ & $\pm 0.0025$ \\ \hline
Overall experimental error 
                  & $\pm 0.0056$ & $\pm 0.0056$ & $\pm 0.0042$ 
                  & $\pm 0.0059$ & $\pm 0.0060$ \\ \hline
Overall theoretical error 
                  & $\pm 0.0050$ & $\pm 0.0034$ & $\pm 0.0066$ 
                  & $\pm 0.0068$ & $\pm 0.0057$ \\ \hline
$\chi^{2}$/d.o.f. & 8.2 / 9  & 5.7 / 13 & 7.9 / 11 & 5.6 / 12 & 4.9 / 8\\
\hline \hline
$\as$(172 \GeV)    & 0.1109 & 0.1099 & 0.1071 & 0.1020 & 0.1121 \\\hline
Statistical error & $\pm 0.0055$ & $\pm 0.0050$ & $\pm 0.0043$ 
                  & $\pm 0.0039$ & $\pm 0.0064$ \\
Systematic error  & $\pm 0.0026$ & $\pm 0.0016$ & $\pm 0.0044$ 
                  & $\pm 0.0022$ & $\pm 0.0024$ \\ \hline
Overall experimental error 
                  & $\pm 0.0061$ & $\pm 0.0052$ & $\pm 0.0062$ 
                  & $\pm 0.0045$ & $\pm 0.0068$ \\ \hline
Overall theoretical error 
                  & $\pm 0.0064$ & $\pm 0.0033$ & $\pm 0.0060$ 
                  & $\pm 0.0065$ & $\pm 0.0057$ \\ \hline
$\chi^{2}$/d.o.f. & 2.8 / 8 & 8.4 / 13 & 7.8 / 12 & 8.4 / 13 & 3.2 / 8 \\
\hline \hline
$\as$(183 \GeV)    & 0.1132 & 0.1075 & 0.1112 & 0.1036 & 0.1081 \\\hline
Statistical error & $\pm 0.0023$ & $\pm 0.0022$ & $\pm 0.0017$ 
                  & $\pm 0.0015$ & $\pm 0.0028$ \\
Systematic error  & $\pm 0.0012$ & $\pm 0.0011$ & $\pm 0.0013$ 
                  & $\pm 0.0006$ & $\pm 0.0010$ \\ \hline
Overall experimental error 
                  & $\pm 0.0026$ & $\pm 0.0025$ & $\pm 0.0021$ 
                  & $\pm 0.0016$ & $\pm 0.0029$ \\ \hline
Overall theoretical error 
                  & $\pm 0.0054$ & $\pm 0.0038$ & $\pm 0.0060$ 
                  & $\pm 0.0071$ & $\pm 0.0054$ \\ \hline
$\chi^{2}$/d.o.f. & 4.2 / 11 & 6.4 / 13 & 15.9 / 12 & 6.3 / 13 & 5.2 / 8 \\
\hline \hline
$\as$(189 \GeV)    & 0.1168 & 0.1108 & 0.1114 & 0.1033 & 0.1118 \\\hline
Statistical error & $\pm 0.0014$ & $\pm 0.0013$ & $\pm 0.0011$ 
                  & $\pm 0.0010$ & $\pm 0.0018$ \\
Systematic error  & $\pm 0.0012$ & $\pm 0.0010$ & $\pm 0.0014$ 
                  & $\pm 0.0012$ & $\pm 0.0014$ \\ \hline
Overall experimental error 
                  & $\pm 0.0018$ & $\pm 0.0016$ & $\pm 0.0018$ 
                  & $\pm 0.0016$ & $\pm 0.0023$ \\ \hline
Overall theoretical error 
                  & $\pm 0.0057$ & $\pm 0.0033$ & $\pm 0.0067$ 
                  & $\pm 0.0078$ & $\pm 0.0055$ \\ \hline
$\chi^{2}$/d.o.f. & 4.4 / 11 & 8.2 / 13 & 28.0 / 12 & 10.6 / 13 & 5.7 / 8 \\
\hline \end{tabular}}
\end{center}
\caption[] {$\as$ measured at $\rs$ = 130, 136, 161, 172, 183 and 189 \GeV\ 
            from fits of the event shape variables to theoretical predictions 
            with combined fixed order and resummed calculations. The fit 
            ranges, the estimated experimental and theoretical errors and 
            the fit quality are also given.}
\label{tab:als}
\end{table}

\begin{table}[htbp]
\begin{center}
\begin{tabular}{|c|c|c|} \hline
$\rs$ (\GeV) & $\as$ (from $T$, $\rho$, $\bt$, $\bw$)
            & $\as$ (from $T$, $\rho$, $\bt$, $\bw$, $C$) \\ \hline
30$-$50 & 0.1400 $\pm$ 0.0056 $\pm$ 0.0107 &                \\
50$-$60 & 0.1260 $\pm$ 0.0073 $\pm$ 0.0088 &                \\
60$-$70 & 0.1340 $\pm$ 0.0060 $\pm$ 0.0087 &                \\
70$-$80 & 0.1210 $\pm$ 0.0064 $\pm$ 0.0082 &                \\
80$-$84 & 0.1200 $\pm$ 0.0057 $\pm$ 0.0089 &                \\
84$-$86 & 0.1160 $\pm$ 0.0061 $\pm$ 0.0082 &                \\
 91.2   & 0.1221 $\pm$ 0.0020 $\pm$ 0.0066 &                \\
 130 & 0.1122 $\pm$ 0.0038 $\pm$ 0.0060 & 0.1128 $\pm$ 0.0038 $\pm$ 0.0063\\
 136 & 0.1116 $\pm$ 0.0042 $\pm$ 0.0060 & 0.1111 $\pm$ 0.0043 $\pm$ 0.0061\\
 161 & 0.1041 $\pm$ 0.0052 $\pm$ 0.0054 & 0.1041 $\pm$ 0.0054 $\pm$ 0.0054\\
 172 & 0.1075 $\pm$ 0.0054 $\pm$ 0.0056 & 0.1084 $\pm$ 0.0056 $\pm$ 0.0055\\
 183 & 0.1089 $\pm$ 0.0022 $\pm$ 0.0056 & 0.1088 $\pm$ 0.0023 $\pm$ 0.0055\\
 189 & 0.1106 $\pm$ 0.0017 $\pm$ 0.0058 & 0.1105 $\pm$ 0.0018 $\pm$ 0.0058\\
\hline\end{tabular}\end{center}
\caption[]{Summary of $\as$ values as determined from event shape variables
           at different centre-of-mass energies. The $\as$ values for $\rs$ 
           $\leq$ $\MZ$ were determined \cite{l3qcd91,l3qqg} only from four 
           event shape variables for which analytical calculations were 
           available at that time.}
\label{tab:alscomb}
\end{table}

\clearpage

\end{document}